\RequirePackage{fix-cm}
\documentclass[twocolumn]{svjour3}          
\def\makeheadbox{\relax}
\smartqed  
\usepackage{graphicx}
\usepackage{enumitem}   
\usepackage{xfrac}
\usepackage{epsfig}
\usepackage{booktabs} 
\usepackage{tikz}
\usepackage{color}
\usepackage{multirow}
\usepackage{graphics}
\usepackage{epsfig}
\usepackage{amssymb,amsmath}
\usepackage{url}
\usepackage{float}
\usepackage{setspace}
\usepackage{parskip}
\usepackage{lastpage}
\usepackage{fancyhdr}

\usepackage[nottoc]{tocbibind}

\usepackage{ragged2e}

\graphicspath{ {figs/} }

\begin{document}

\title{Structuring Communities for Sharing Human Digital Memories in a Social P2P Network
}

\titlerunning{Digital memories in a social P2P network}        

\author{Haseeb~Ur~Rahman         \and
        Madjid~Merabti  \and
				David~Llewellyn-Jones   \and
				Sud~Sudirman		\and
				Anwar~Ghani 
} 

\authorrunning{Rahman et al.} 

\institute{A. Ghani* \at
              Department of Computer Science \& Software Engineering\\ International Islamic University Islamabad, Pakistan\\
              Tel.: +92-51-9019725\\
              \email{anwar.ghani@iiu.edu.pk}       
           \and
           H. Rahman \at
              University of Malakand, Pakistan
					\and
           M. Merabti \at
							University of Sharjah, Sharjah, UAE
				  \and
           D. Llewellyn-Jones \at
							University of Cambridge, United Kingdom
					\and
           S. Sudirman \at
							Liverpool John Moores University United Kingdom
}

\date{Received: date / Accepted: date}

\maketitle

\begin{abstract}
A community is sub-network inside P2P networks that partition the network into groups of similar peers to improve performance by reducing network traffic and high search query success rate. Large communities are common in online social networks than traditional file-sharing P2P networks because many people capture huge amounts of data through their lives. This increases the number of hosts bearing similar data in the network and hence increases the size of communities. This article presents a Memory Thread-based Communities for our Entity-based social P2P network that partition the network into groups of peers sharing data belonging to an entity--person, place, object or interest, having its own digital memory or be a part another memory. These connected peers having further similarities by organizing the network using linear orderings. A Memory-Thread is the collection of digital memories having a common reference key and organized according to some form of correlation. The simulation results show an increase in network performance for the proposed scheme along with a decrease in network overhead and higher query success rate compared to other similar schemes. The network maintains its performance even while the network traffic and size increase.

\end{abstract}

\section{Introduction} 
\label{intro}
Human mind stores information in an encoded form~\cite{tpds01,tpds02}. This is a biological event, and the encoding process is carried out through various signals generated in the brain. The signals are generated as a result of human perceptions and experiences of its surroundings, through their senses, in the form of objects, places, people, events, emotions and so on. These stimuli such as people, places, etc. are the cues to store and recall human memories. The cues are also interconnected, such that one cue can lead to the storage or recall of a different memory. For example, while people can generally intentionally recall specific memories about friends, family, places, events, and so on, seeing a related scene will often result in the (involuntary) recollection of other memories~\cite{h01}. An example might be the revisiting of a location a second time reminding us of the friends with whom together we visited the place before. The relationships between memories can, therefore, be important for their recollection, and we use this fact to organize human life digital memories in the form of Memory Threads. This leads on to the organization of our digital memories into Memory-thread Based Communities (MTC).

The decrease in the cost of personal devices such as mobile phones, digital cameras and so on has allowed people to capture and store their life memories in a digital form far more easily than ever before. Each digital memory is collected by someone at a time, place or an event and involving different people, objects and so on. We consider these real-world objects, interests, places, and people and refer to them as Entities. An entity also represents a co-relation between these digital memories. For example, a person would represent a connection between all the digital memories in the form of pictures, audio captures, videos, and so on, which are captured, collected or stored by that person or for them throughout their life at different times, places, events and so on. We believe that the correlations of these digital memories can allow them to be organized in a meaningful way. Organized digital memories help in presenting the purpose of the data for which the memories were captured and recalling them in future in an intended way.

An entity can be considered to be anything which has its own digital memories captured or stored by itself or can form part of the digital memories of others. Entities present within a set of memories can be identified by various tools such as a ‘Memory for Life’ (M4L) system~\cite{tpds03}. An M4L system can analyze and annotate data, and detect entities which exist as part of it. For example, indoor or outdoor images, people, places, objects in an image and so on might all be considered as entities. Information about entities is often stored in the form of metadata which can further be used to organize the original data. 

In order to take advantage of the connections between digital memories, this article proposes the use of memory threads to organize sets of digital memories. A memory thread is formed for an entity that represents the events or sequence of events that occurred for that entity. Digital memories in a memory thread are selected according to a selection criterion, which is usually the existence of the entity. They can then be organized according to an indexing criterion. The digital memories in a memory thread are organized in a way that expresses information about the entity. Multiple indexing criteria can be used to organize the digital memories into a multi-dimensional space such that applying a single or combination of criteria can retrieve the required digital memory or memories. For example, the time or period within a person’s life that memory was captured might constitute a criterion for storing and recalling that person’s memories. In this example, the person, as an entity, is the selection criterion and the age of the person constitutes the indexing criterion. Almost every digital memory is stored with the time at which the digital memories were captured. Therefore, we can organize digital memories based on the history of the entity they relate to, which is likely to be useful since history is one of the ways to express information about objects, places and so on. If we organize digital memories according to the history of an object, it naturally forms a linearly structured memory thread.

In our previous work~\cite{tpds04}, different challenges has been identified for sharing human digital memories such as network structure, data privacy, searching, etc.~\cite{tpds31,tpds32,tpds33} and have proposed an Entity-Based Social Peer-to-Peer (P2P) Network (ESP2PN)~\cite{tpds05} where the communities are formed according to entities in digital memories~\cite{h10}. This operates on the analyzed and annotated digital memories captured by memory for life system. We have assumed the digital memories of each person are stored on and shared through their own peer or personal social network~\cite{h09}. Considering the importance of the underlying network structure can have a significant effect on the performance of the network~\cite{tpds25,tpds26,tpds27,tpds28,tpds29}, a Memory Thread-based Communities approach has been proposed to address this issue. An MTC reflects memory threads in a similar way to the method we use to structure peers in these memory thread-based Communities. An MTC is formed by peers that share data for an entity such that each peer in it connects according to the position defined, based on the digital memories it shares, by the indexing criterion. Memory thread-based communities, therefore, bring together similar peers together, with closeness also being related to the ordering of the indexing criteria. For example, if a person is sharing the digital memories of his school years good places for them to initially connect to are the peers storing digital memories relating to the same School and period, which are likely to be those belonging to classmates. We believe that structuring the community according to certain criteria gives topological awareness to a peer for routing search queries~\cite{tpds26,tpds34}, which will result in improving the efficiency of search. It also makes these communities more scalable because a peer connects only to a few neighboring peers – those that are most similar based on the criteria – as part of a larger community.  Moreover, the indexing criteria will give certain information to a user for browsing the social P2P network.

The general contribution of the this article can be listed as follows:
\begin{enumerate}
  \item Human digital memories exists with certain patterns that allows them to be organized meaningfully. The first contribution is to proposed a novel Memory-thread based communities that are organized according to the patterns in the digital memories.
	\item To present Memory Thread-based Communities for Entity-based social P2P network that partition the network into groups of peers sharing data belonging to an entity to handle the challenge of reducing the number of hosts bearing similar data as well as reduce the size of the community. As a result large communities are organized meaningfully according to a common reference key and a correlation among them.
	\item To improve the network performance by decreasing the network overhead and increasing the query success rate compared to other similar schemes. 
	\item To maintain the network and its performance irrespective of the increase in network traffic and size of the network. 
	\item To maintain scalability of services irrespective of the networks size and traffic 
\end{enumerate}

In the remainder of this paper, we will elaborate on these ideas further. Section~\ref{relwrk} describes existing work and the arguments as to why existing architectures are not suitable in our case. Section~\ref{memTh} explains the idea of memory threads, its types and how they can be used to organize digital memories. Section~\ref{memThrcomm} explains how the communities are formed according to memory threads. Here we explain how various types of memory thread can be reflected in the structure of the network and the joining of a peer to a memory thread-based community. We provide details of our simulated memory thread-based community system and compare it with some of the existing file sharing P2P networks aimed at tackling similar digital memory storage issues. The results are explained in Section~\ref{simSetup}. The article has been concluded in Section~\ref{conclu} with a look at future work.

\section{Related Work}
\label{relwrk}

Methods for organizing digital memories using computing devices have been considered since the birth of the computer, starting with Vannevar Bush in 1945~\cite{tpds06} in his famous article ``As We, May Think'' in the form of a machine called the Memex. According to Bush, ``A memex is a device in which individual stores all his books, records, and communications, and which is mechanized so that it may be consulted with exceeding speed and flexibility. It is an enlarged intimate supplement to his memory''. At the push of a button, all the data of a person should be retrieved by the machine in a short time . Gorden Bell’s MyLifeBits~\cite{tpds07} aimed to provide a realization of the Memex utilizing the tools, increased processing capabilities and relatively-speaking large digital storage capacities that have become available more recently. The software of MyLifeBits has the ability to store images, links, text,  videos, etc. in a database and allows manual annotation of the material. The area has subsequently been further extended to collecting digital memories and various tools and techniques have been developed for this e.g. Eyetape. Jim Gemmell et al.~\cite{tpds08} describe the four principles that were applied when designing MyLifeBits. First, ``strict hierarchy should not be imposed on data organization''. Second, ``many visualizations of their life bits were desirable to help understand what they would be looking at''. Third, ``the value of non-text media is dependent on annotations''. Fourth, ``authoring tools create two-way links to media that they included with new media''. Several other areas were identified and a single platform was required to develop tools and techniques for collecting, storing, organizing, sharing and generating meaningful information from digital memories. In this effort, developing an M4L system was accepted as a third grand challenge (GC3) by the Engineering and Physical Science Council (EPSRC)~\cite{tpds09} in the UK in the year 2008~\cite{tpds10}. Azizan et al.~\cite{tpds11} are currently working on a prototype of a Human Life Memory system for collecting, storing and organizing different life events, and for identifying interesting events referred to as ``Serendipitous Moments'', as well as discussing sharing via P2P networks. In our previous work, we described the challenges for sharing human life digital memories~\cite{tpds04} and proposed an Entity-based social P2P network~\cite{tpds05} based on the challenges for sharing memories. To share digital memories, it is important to properly organize the network so that data of the entities is not lost. The following are a few approaches which use the data contents to organize and search peers and data in the network.

Unstructured P2P networks \cite{h02,tpds12,tpds31,tpds32,tpds33,tpds34}, also called pure P2P networks, connect peers in a 'random' manner. Each peer has equal responsibility for routing messages and providing services. Since there is no central system that controls and manages the network, it is the responsibility of each peer to keep a record of its neighbor peers and resources. Random walk~\cite{h08} and Flooding are the most commonly used searching techniques in unstructured networks \cite{tpds13,tpds32,tpds34}. Unstructured networks are less scalable, produce high network overhead and have lower search precision (due to queries being directed to irrelevant peers) compared to their structured counterparts \cite{tpds14,tpds15,h11,h12}. Search precision also depends upon the number of hops a query travels. If the number of hops a query travels is higher, then the successful query rate increases while at the same time generating more network overhead and vice versa.

Upadrashta et al.~\cite{tpds16} utilize the in-network experience of a peer. Peers analyze queries and find the interest of a peer from the search queries that are received from other peers. In this way, every peer keeps (stores) information about other peers, leading to the formation of virtual communities. Upon the arrival of a search query, it has been analyzed and then forwarded to peers with similar interests reflected in the search query. An inexperienced peer has less knowledge, making it harder for it to find content stored by others, which can be a problem for newly joined peers. In Semantic Overlay Networks (SON)~\cite{tpds17}, peers that are semantically similar are grouped into a single cluster. Maze~\cite{tpds18} is a centralized online social P2P network, which allows people to share their resources, friends list and status. The server is responsible for connecting peers and issuing tickets to peers for security. However, the network cannot operate without the server due to the crucial role it plays in providing connectivity and security. In a similar approach proposed by Modarresi et al.~\cite{tpds19}, a group of peers with similar interests are gathered in a community. Data lookup in Interest-Based Communities (IBC) is performed by sending queries only to members with similar interests. Community-based approaches group similar peers together which allow peers that do not have the required data to be avoided during the search. Such a community certainly brings similar peers together but does not provide any information about the status, characteristics or ‘personality’ of entities. Within these communities, there is no further structure imposed, and peers will connect with other peers without consideration for any other similarity criteria. The exact location of data is therefore not known in advance. When a peer searches for some particular item of data, it sends queries to all members – or up to a certain number of hops – within a community. However, sending queries to all peers creates overhead since peers that have no relevant data will also receive these queries. Lowering the number of hops inside the community reduces the overhead, but decreases the chance of finding the required data.

In IBCs, data is shared according to the interest of the host. If the interest of the host changes then there is the possibility that their associated peer will leave the community and the data being shared by the peer will not be accessible anymore. Contrary to IBCs, we consider the actual data being shared by the host for creating communities and connecting peers inside each community; because as long as the data is available for sharing, it will be accessible in the same community. We also believe that our memory thread-based communities will be more stable due to the entities in our Entity-based social P2P network.


\section{Memory Threads}
\label{memTh}
 
Various hardware and software tools, such as Memory for Life systems~\cite{tpds20}, are able to analyse and meaningfully define memories in a digital form. Some systems where automatic analysis by a tool is difficult, also allow manual annotations to be made. MyLifeBits provides a good example of this case~\cite{tpds08}. We refer to the tagged information stored in the form of metadata (either by hardware, software or manually) with a digital memory as a memory key. As described above, people remember their memories based on some reference, such as person, place, event etc. which we call entities, which is then used for recollection. The memory key(s), in the form of entities in data, contain such reference(s) to a digital memory that can be used to recall it is called a reference key. A single or combination of such memory keys forms a reference key. A digital memory must have at least one reference key. A reference key for a piece of data can either be set by the owner or user of the data explicitly, or obtained from a data analysis and annotation tool such as an M4L system.

\begin{figure}
  \centering
  \includegraphics[scale = 0.8]{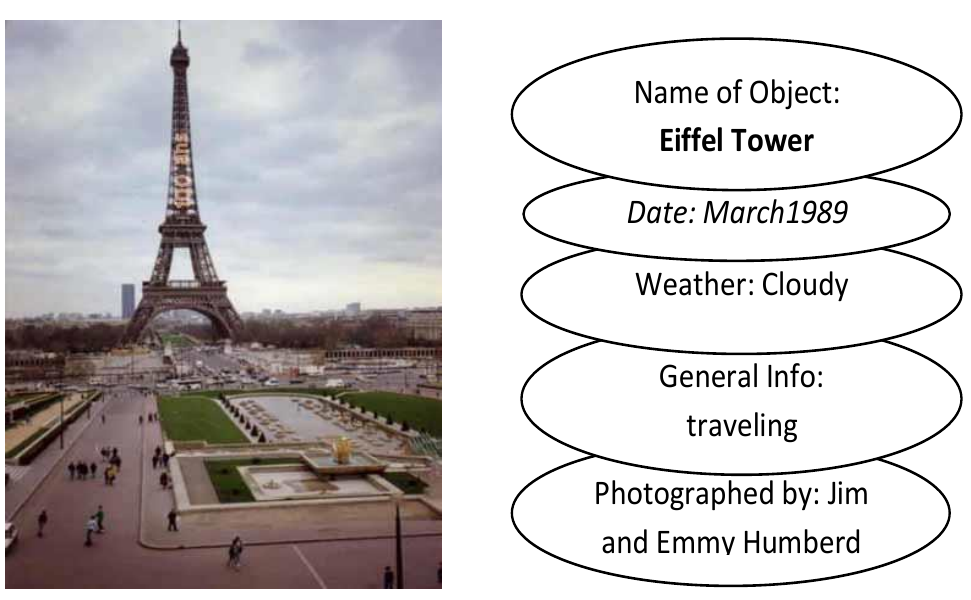}
\caption{A digital memory and its associated memory keys.}
\label{fig:fig1}
\end{figure}

We will explore this idea more comprehensively using an illustrative example (adapted from a real scenario) based on the picture of the Eiffel Tower shown in Fig.~\ref{fig:fig1}~\cite{tpds21}. This picture was captured by Jim and Emmy Humberd in March 1989 using a digital camera in cloudy weather during the when it was refurbished for its 100th anniversary. In this example the type of data (picture), date (March 1989), device (camera), name of entity (Eiffel Tower), weather (cloudy), photographer (Emmy Humberd) and event (2nd year of 100th anniversary) are all memory keys. Reference keys to the memory in Fig.~\ref{fig:fig1}, in the form of the names of the entities involved, can be set as ``Eiffel Tower'' and the photographers. The digital memory not only belongs to the Eiffel Tower but also forms part of the digital memories of the photographers.

Similar digital memories have a common reference key and can be recalled using the same reference. Digital memories of similar references can be put together and are organized according to certain criteria to form a memory thread. In other words, a memory thread is the collection of memories which has a common reference key and is organized by a criterion which structures them in a specific order. The order of the digital memories in a memory thread should be established so that they can be traced by moving from one memory item to another. A memory thread should also be in a form that provides information about the entity for which the memory thread has been formed. For example, time, which is further considered as history of an entity to form memory thread, is often used as a reference to digital memories since it allows their memories to be recollected based on when an event took place.

\subsection{Selection \& Indexing Criteria}
\begin{figure}
  \centering
  \includegraphics[scale = 0.9]{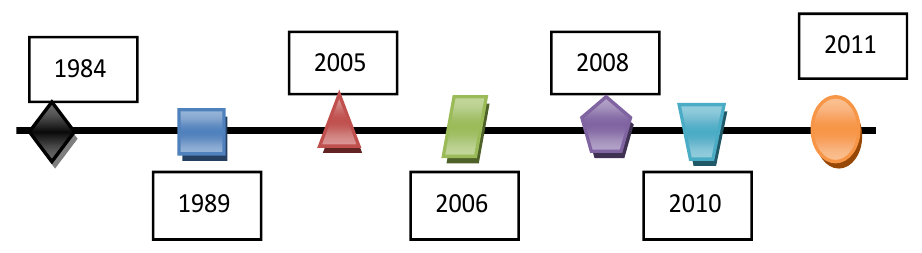}
\caption{A virtual memory thread based on entity (selection criterion) and time (indexing criterion)}
\label{fig:fig2}
\end{figure}

 To form a memory thread, certain criteria are used to select the relevant digital memories that it will be made up of. As we discusses above, there are two types of criteria used to form a thread: selection criteria and indexing criteria. The selection criteria select those digital memories that have a similar reference key as used by people to recall their memories. The indexing criterion organizes the digital memories into a given order. It's, therefore important that every digital memory should have at least one memory key which can be used as an indexing criterion. For example, to form a memory thread for the Eiffel Tower, the selection criterion would be the name of the entity itself (i.e. the Eiffel Tower). However, as a selection criterion this will be invariant across all of the memories, and therefore wouldn't generally be appropriate as an indexing criterion as well (although there are certain situations where this is not the case, such as where the selection criterion takes the form of a continuous range of values). A separate indexing criterion is therefore needed, examples of which could be time, temperature, or any other total ordering. If time is the indexing criteria for the Eiffel Tower thread, then the thread will be formed as the digital memories of the structure at different times, for  example relative to its construction in 1889. A memory thread of the Eiffel Tower is illustrated in Fig.~\ref{fig:fig2}, where each point represents a digital memory along with the time a digital memory is captured. In this example, the memory thread gives us information about the history of the entity. Yet there is no single memory key that allows organizing all types of digital memories under a single criterion, which is why it is not possible to present a specific structure for memory threads: the structure will change according to the indexing criterion that organizes the data. For example, as considered in the above example time can be a memory key for organizing digital memories of the Eiffel Tower. If digital memories for the Eiffel Tower are organized according to time then it will form a linearly structured memory thread, but the structure of the memory thread would be different if we were to choose location as an indexing criterion.

\begin{figure}
  \centering
  \includegraphics[scale = 0.9]{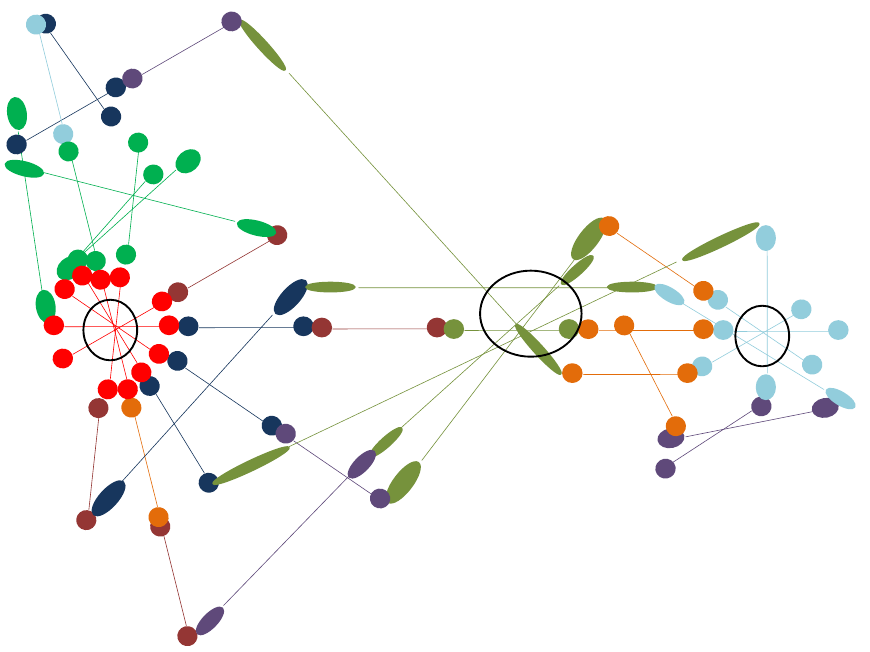}
\caption{A network of extant memory threads}
\label{fig:fig3}
\end{figure}

A memory thread is formed for every entity, which has its own digital memories or is a part of the memories of another entity. A memory thread stretches over many digital memories that relate to a single entity. At a certain place in a memory thread, where the digital memories of different entities have a common reference key, different threads will overlap with one another. We call the overlapping points of different threads networking points, shown in Fig.~\ref{fig:fig3} as circled areas. A memory thread can have a single or many overlapping points. For example, many people, store the digital memories of their visit to the Eiffel Tower as a part of their memory threads. Each of them has their own memory threads, but at the Eiffel Tower their memory threads overlap with each other which become a \textbf{networking point} for them. The networking points connect different memory threads and form a network of memory threads. The networking points allow access to the memory threads of other entities. In our social life, we also have networking points. For example, a conversation becomes friendlier when two people have visited the same place or they belong to the same profession.
\subsection{Extant and Virtual Memory threads}

\begin{figure}
  \centering
  \includegraphics[scale = 0.9]{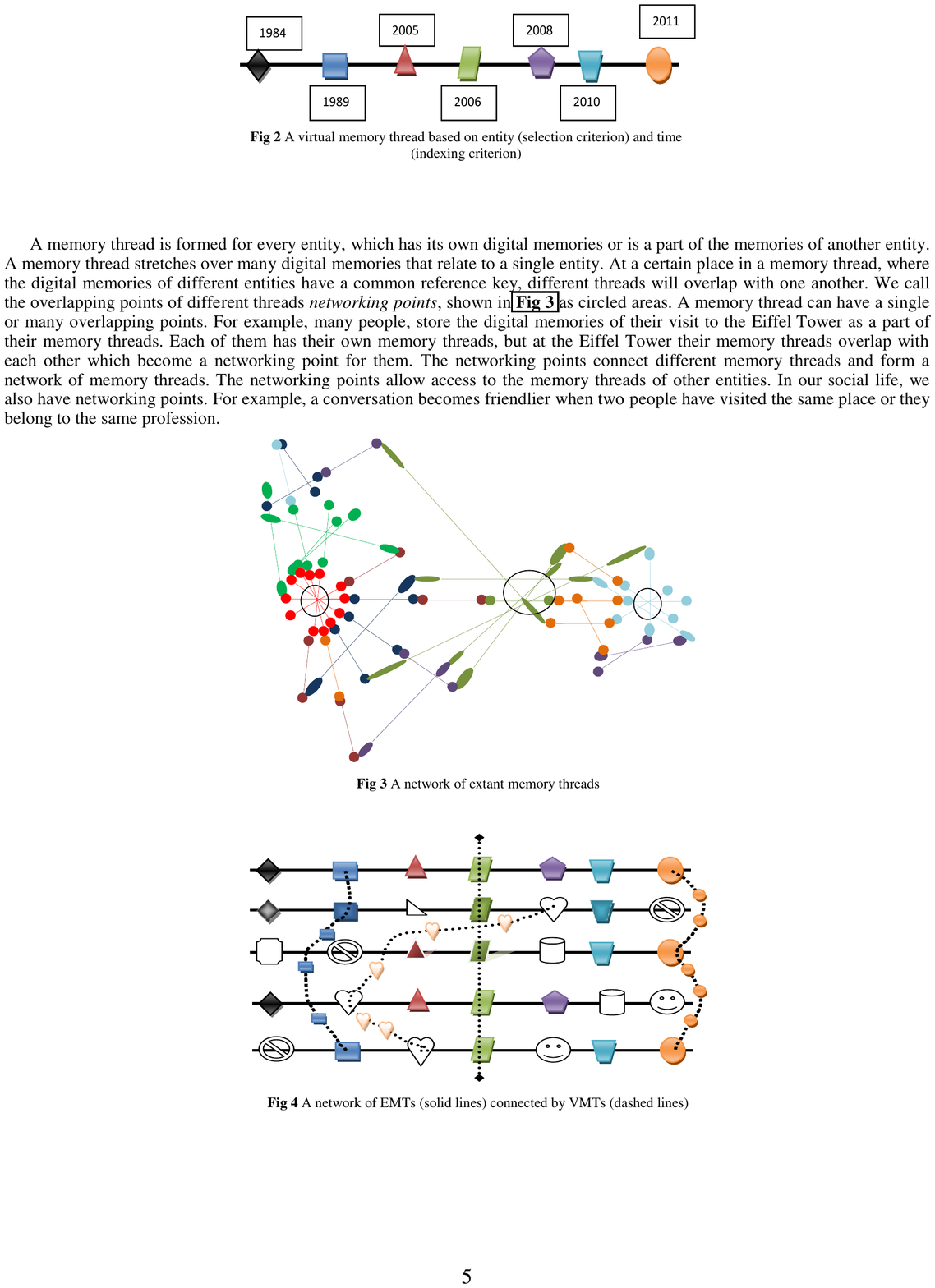}
\caption{A network of EMTs (solid lines) connected by VMTs (dashed lines)}
\label{fig:fig4}
\end{figure}

As defined above, some entities – such as humans – are capable of capturing, storing and maintaining their memory threads using a device such as a computer, mobile phone etc. Such a memory thread would be called an Extant Memory Thread (EMT). The memory thread of an entity whose memories are part of the extant memory threads of other entities, but where the entity cannot itself capture, store or maintain memories, is called a Virtual Memory Thread (VMT). A virtual memory thread is formed at the networking points of the EMTs as shown in Fig.~\ref{fig:fig3}, because these will be the memories which belong to/involve those entities which either do not own or cannot store, capture or maintain their memories. This can be counter-intuitive, since we wouldn't normally refer to memories of entities that are not capable of having memories. However, in our case it's convenient to project memories onto inanimate objects for the purposes of generalization and improving the effectiveness of the network. Each entity can then find the memories of other entities by following its extant memory threads to a networking point that can lead to the virtual memory threads of the entities it is interested in. Fig.~\ref{fig:fig4} illustrates a selection of EMTs connected by VMTs. 

\section{The Proposed: Memory Thread-Based P2P Communities}
\label{memThrcomm}

A detailed discussions on Memory Thread-based P2P community, How EMT and VMT based communities are formed in MTCs, structure of an MTCs and how peer join the communities are given further below.

\subsection{Memory Thread-based P2P Communities}

The network structure of our Memory Thread-based P2P Communities for Entity-based social Peer-to-Peer (P2P) networks ~\cite{tpds05} is based on the idea of memory threads described in the previous section. In our approach, we maintain memory threads for each entity across the network. A memory thread is reflected in the network structure in order to organize the peers in the form of a Memory Thread-based Community (MTC). A peer in an MTC represents a digital memory in a memory thread formed in the network. Since, a peer can store many digital memories either for a single or many entities, which allows it to become a member of one or more than one MTCs. Similar to memory threads, a selection criterion, which is an entity, is used to form an MTC and an indexing criterion defines a structure or an order for the community. In our network, a reference key for a digital memory can be set in a number of ways. It can be set by the owner of the data explicitly, set automatically by an M4L system, or obtained by aggregating search queries received for a single or a group of memory keys. As described earlier, an entity can have an extant memory thread or a virtual memory thread; these are also used to organize peers in our network. An extant memory thread is maintained on a single peer and a virtual memory thread across the network. In other words, a virtual memory thread spans many extant memory threads stored at different peers, which allows to connect peers in network in the form of MTCs.

\subsection{Types of Communities}

Communities has been classified into two categories and briefly explained in the following sub section.

\subsubsection{Extant Memory Threads based Communities} 

A community of peers of Extant Memory Thread (EMT) is formed by or for an entity from its digital memories. As people prefer to be the sole owner of their own data, therefore, an EMT is stored on a single peer or in the personal social P2P network of an entity because data in an EMT contains only a user's own captured digital memories. To create an extant memory thread, an entity specifies selection and indexing criteria and simply binds their existing digital memories to the thread (or adds them to the thread when a new digital memory is captured) on his own peer or personal social P2P network. Note that entities can have a single or more extant memory threads and that a digital memory can be part more than one extant memory thread at a time (but must be part of at least one). Other peers in the network join an EMT either by invitation or as members in a virtual memory thread. An EMT is simply a memory thread stored on a single peer to which other peers in the network are attached.

\subsubsection{Virtual Memory Threads based communities}

Creating a P2P community of Virtual Memory Thread (VMT) is less straight forward, due to its decentralized nature. As we discussed above, there are entities that cannot capture, store or maintain their own digital memories and their ‘memories’ therefore all – by necessity – form part of the digital memories of other entities. Also, for entities which can capture or store their memories there are nonetheless some digital memories that relate to them but are not owned by them and cannot therefore become a part of their EMTs. Such digital memories will be part of EMTs of other entities and could possibly be stored at more than one peer. A VMT connects those peers in a community that store data belonging to a single entity. Memory thread-based communities are actually formed using VMTs. The following paragraphs describe the process of forming and joining a VMT in the network.

To form a virtual memory thread, a peer broadcasts a request for the digital memories of an entity. The request contains the selection criterion for the memory thread. Those peers that have data matching the criteria of the thread reply with a notification highlighting the availability of such data. The reply message contains information about the stored data in their EMT which can become a part of the VMT. The sender receives the replies from all the peers and starts a new thread-based community by making a list of those peers which claim to have relevant data. A suitable indexing criterion is then applied to these digital memories to structure the list of peers. The list contains the addresses of the peers structured according to the indexing criteria. The list is sent to all peers that replied with relevant data and the peers become a part of the newly formed VMT.

\subsection{Joining Memory Thread-Based Community}

A memory thread-based community is formed by connecting those peers which have common reference keys for similar digital memories. Peers in the MTC are arranged according to the indexing criterion in a specific order as described above, and this ordering also defines the structure of the community. New peers can join an existing community, as soon as they are discovered or new digital memories are added by entities in the network.

If a peer wants to join a memory thread, it follows the indexing criteria to find its place in the thread. A peer finds the first peer by broadcasting in the network as described earlier; once found it connects with it. Each peer in the thread retains information about its neighbor peers either side in this thread. A new peer can therefore find its location in the thread by sequentially sending messages to each peer along the thread. When a peer finds its suitable location in the thread, it then stores its two or three hop neighbors on both sides in the thread. The purpose of connecting with two or three hop neighbours instead of one is to avoid partitioning occurring in the event that a peer leaves the network unexpectedly. In this case the next hop neighbour will be connected which will avoid the thread becoming split. There will be no dedicated peer responsible for maintaining a memory thread; if any peer – whether the peer that started the thread or one that joined later – leaves a thread, it will not disconnect the thread since each peer has equal responsibilities in maintaining the thread.

\subsection{The Structure of Memory Thread-Based Communities}

\begin{figure}[h]
\centering
	\includegraphics[scale = 0.8]{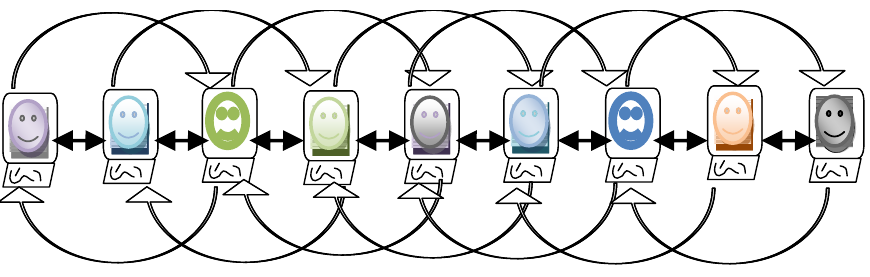}
	\caption{Memory thread-based community with 2-hop connection.}
	\label{fig:fig5}
\end{figure}

The purpose of memory thread-based communities is to organize the network properly. For this purpose, a characteristic of an entity, stored in the form of metadata, can be used to form a memory thread in order to organize the community for an entity. To structure our network for achieving good network performance, we consider the use of the history of an entity to organize its digital memories within the network. One important reason for choosing this as an indexing criterion is that the time at which a digital memory is captured is usually stored with the memory, and is therefore widely available. For this indexing criterion the peers in our network will be organized linearly such that each peer stores the address of its immediate one and two hop neighbors, as shown in Fig.~\ref{fig:fig5}.

In our network, two important scenarios have been considered that have an effect on the average path length and robustness of the network. First, a single peer may store files that result in its appearing multiple times and at different places in a single memory thread. These types of peers can provide shortcuts while routing through the network. In the second scenario, multiple peers store duplicate files. The indexing criteria place all of these multiple files into a single location within the memory thread. So if any peer fails or leaves the network then another peer storing the same (or a very similar) file will remain available, thereby increasing the robustness of the network.

\begin{figure}[h]
\centering
	\includegraphics[scale = 0.8]{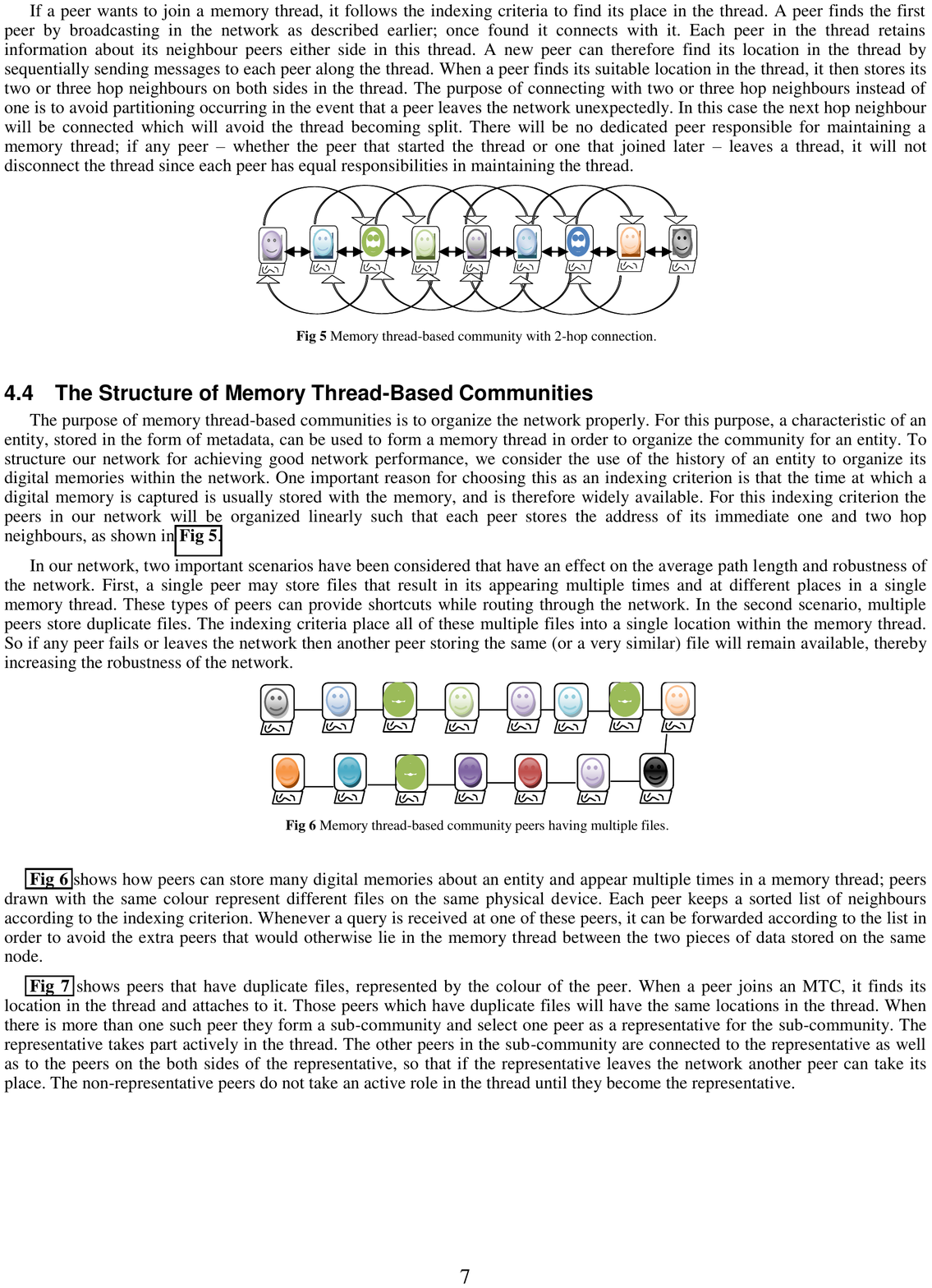}
	\caption{Memory thread-based community peers having multiple files.}
	\label{fig:fig6}
\end{figure}

Fig.~\ref{fig:fig6} shows how peers can store many digital memories about an entity and appear multiple times in a memory thread; peers drawn with the same colour represent different files on the same physical device. Each peer keeps a sorted list of neighbours according to the indexing criterion. Whenever a query is received at one of these peers, it can be forwarded according to the list in order to avoid the extra peers that would otherwise lie in the memory thread between the two pieces of data stored on the same node.

\begin{figure}[h]
\centering
	\includegraphics[scale = 0.8]{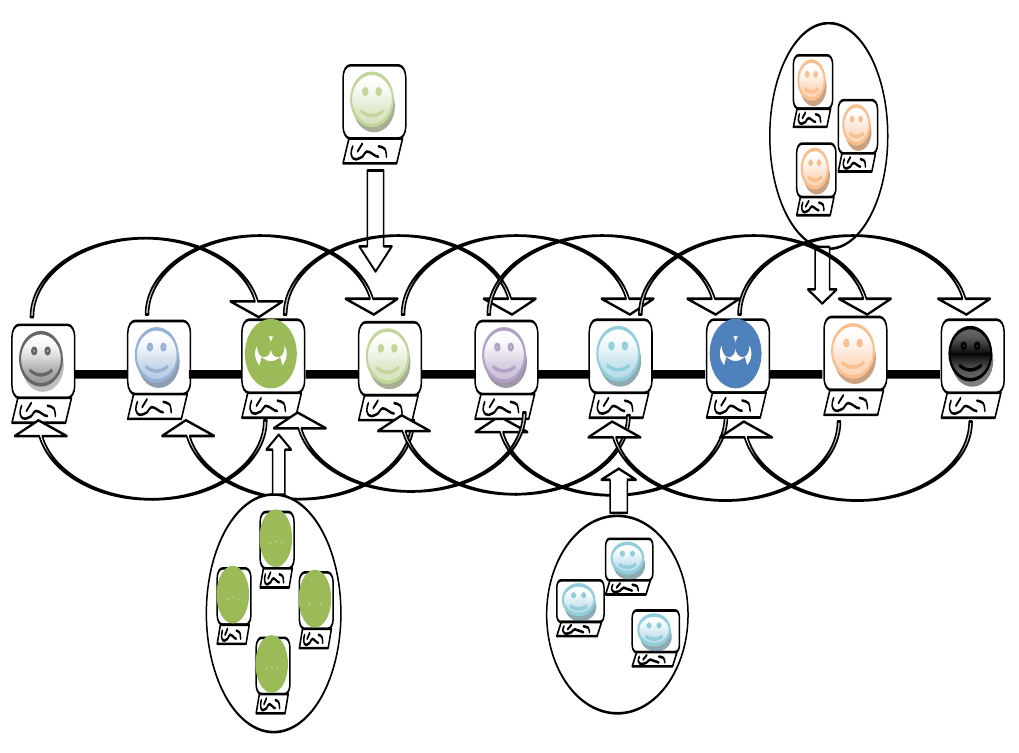}
	\caption{Memory thread-based community with duplicate digital memories.}
	\label{fig:fig7}
\end{figure}

Fig.~\ref{fig:fig7} shows peers that have duplicate files, represented by the colour of the peer. When a peer joins an MTC, it finds its location in the thread and attaches to it. Those peers which have duplicate files will have the same locations in the thread. When there is more than one such peer they form a sub-community and select one peer as a representative for the sub-community. The representative takes part actively in the thread. The other peers in the sub-community are connected to the representative as well as to the peers on the both sides of the representative, so that if the representative leaves the network another peer can take its place. The non-representative peers do not take an active role in the thread until they become the representative.

The following figures, Fig.~\ref{fig:fig8} and Fig.~\ref{fig:fig9}, show a snapshot of our MTCs, for the simulated network that we will discuss in Section 5, generated by Pajek [22]. Fig.~\ref{fig:fig8} shows all the peers in a circular form. The edges in the diagram indicate shorts paths running across the circle of peers. Fig.~\ref{fig:fig9} is the network diagram of the peers in Fig.~\ref{fig:fig8}, without imposing any circular arrangement.

\begin{figure}[h]
\centering
	\includegraphics[scale = 0.8]{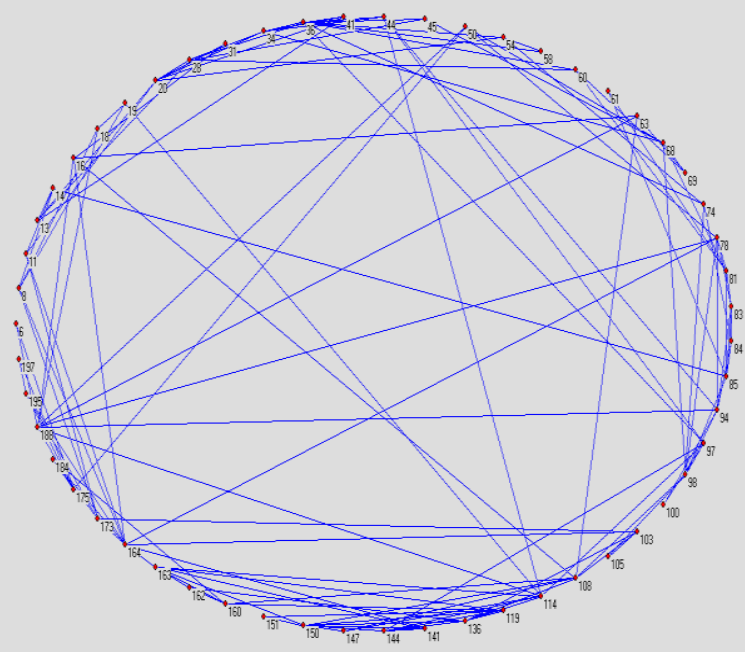}
	\caption{Memory thread-based community showing short paths}
	\label{fig:fig8}
\end{figure}

\begin{figure}[h]
\centering
	\includegraphics[scale = 0.8]{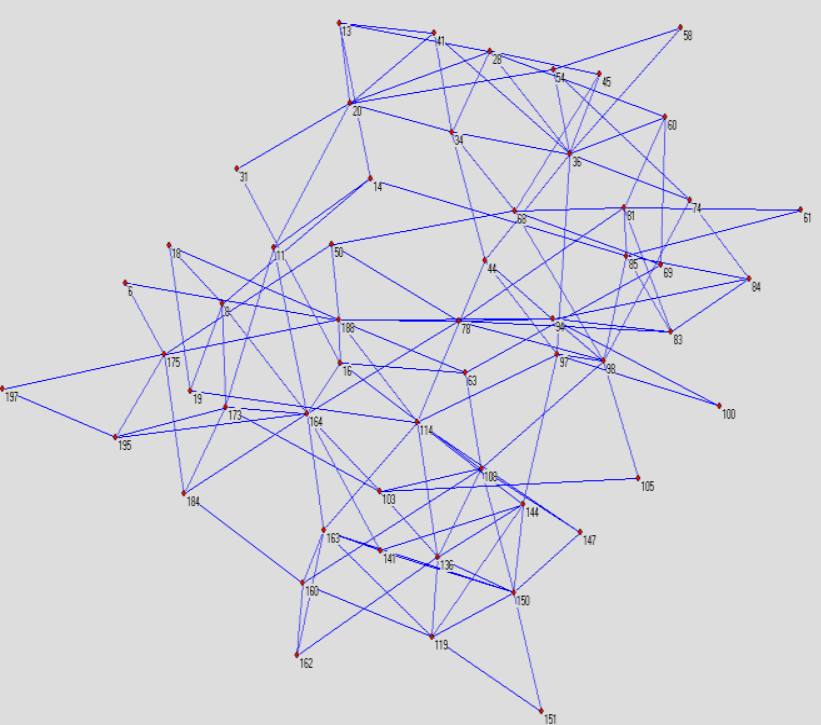}
	\caption{Network of peers of the Memory thread-based community}
	\label{fig:fig9}
\end{figure}

\section{Simulation Results}
\label{simSetup}

In this section we describe our simulated MTC network and compare it to existing methods. In particular we compare it in terms of performance to interest-based communities (IBC) and unstructured networks. The simulator has been adapted from that of Modarresi et al.~\cite{tpds23} and is used for simulating social networks and semantic social networks. 

\subsection{Simulation setup}

\begin{figure}
\centering
	\includegraphics[scale = 0.75]{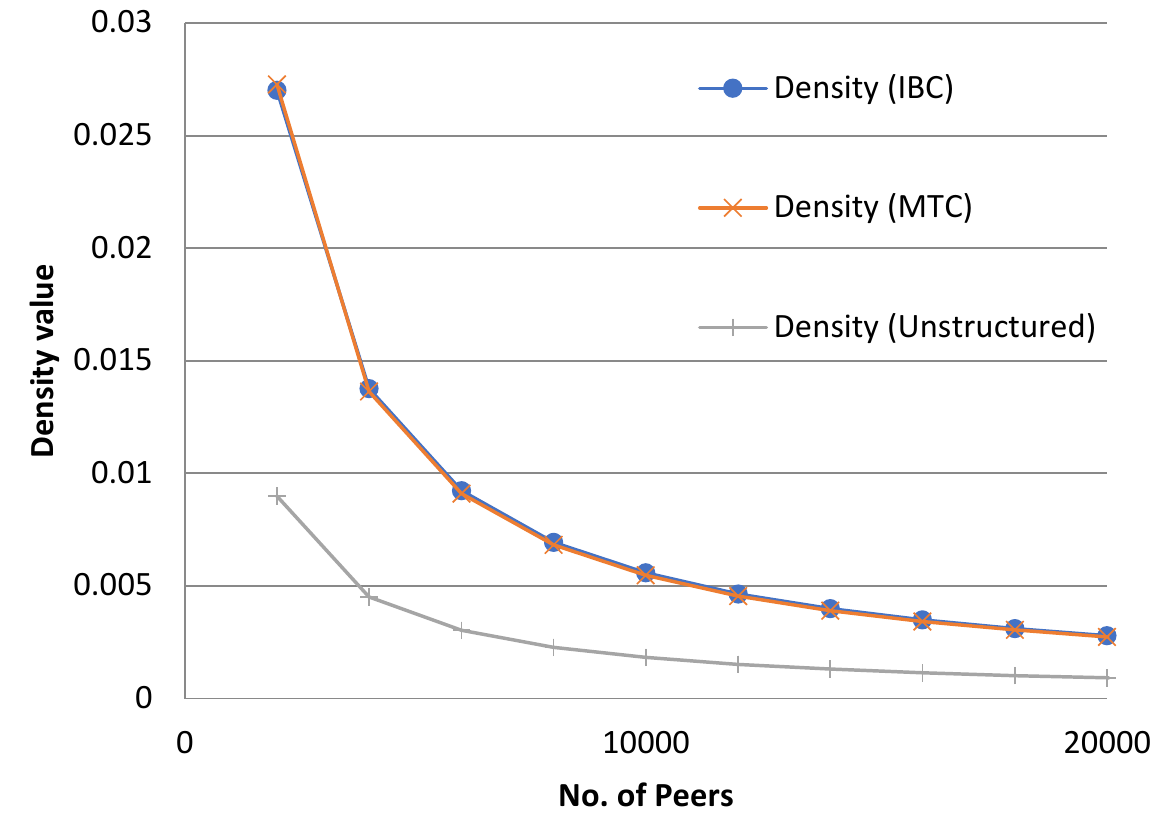}
	\caption{Graph density of the networks}
	\label{fig:fig10}
\end{figure}

The simulation environment was set according to the static input parameters so that the network structure follows a power law distribution. This is to reflect the behaviour expected of an online social network~\cite{tpds24}. The total number of interest-based communities in the simulation was set to seven and sub-communities are linearly distributed in each community. Each peer is a member of three communities at a time and files are distributed linearly throughout the network. Each peer has an upper limit set for the number of connections it can support, determined by a neighbour distribution function that obeys a power law distribution with scale value greater than two. Network connections are bi-directional and the simulation time is 5000 simulation seconds. Peers in the unstructured network are connected randomly by selecting a peer and connecting it with a random number of peers until it reaches its upper limit of connections. The files are also distributed randomly. The graph density is calculated according to the formula $(2\times E) / (V\times (V - 1))$ where E is total number of edges and V is total number of vertices. For each of the above networks this is shown in Fig.~\ref{fig:fig10}.

Flooding has been used as the search technique for all of the networks and the maximum number of hops a query can travels is capped at three. The total number of connections in each network is approximately the same, as given by the graph density in Fig.~\ref{fig:fig10}. All results are given as the average value over fifteen experimental runs. The outputs measured are the rate of successful queries, the resulting network overhead and the number of hops a query travels if successful.

To measure the output, the input conditions varied to test the performance of the network are the following.
\begin{enumerate}
	\item Query distribution: Queries in the network are created according to a linear distribution function. The distribution function which is used to distribute the events, and to send new queries in the network within the 5000s time limit is:
\begin{equation}
	T_{n+1} = T_n + (b\pm m)
\label{eq1}
\end{equation}
where Tn is the time of the last event, Tn+1 the time of the next event to be set, b is a base value for the average delay and m is a modifier. The base b for the function takes the values 200s, 100s, 60s, 40s, 30s, 25s, 20s and 15s and the modifier m takes the value 40s. The base value represents the time interval between sending new queries in the network. The modifier value is used by the distribution function to vary the interval. This affects the amount of traffic in the network: the lower the base value, the greater the network traffic and vice versa. By changing network traffic, we test the successful queries in the network and the amount of overhead created. For these runs the network size is set at a constant 5000 peers.
\item Number of peers: The number of peers in the network has also been varied to test the performance of the network with different node densities. The traffic in the network is generated according to the linear distribution $T_{n+1} = Tn + (b\pm m)$ as before, but this time with a constant base value b of 100 and constant modifier m of 40 throughout all experiments. The successful queries in the network and the level of overhead created were recorded while increasing the network size, which ranges from 2000 to 20000 peers.
\end{enumerate}

\subsection{Results and comparison}

\begin{figure}[h]
\centering
	\includegraphics[scale = 0.8]{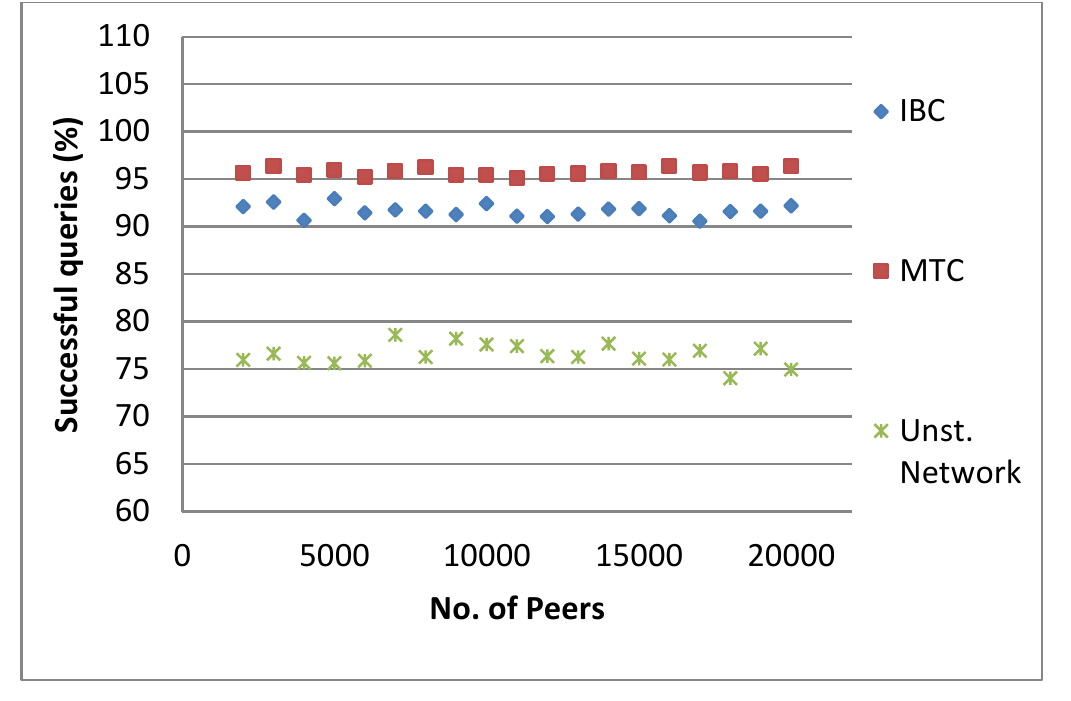}
	\caption{Percentage successful queries as a result of increasing network size}
	\label{fig:fig11}
\end{figure}

\begin{figure}[h]
\centering
	\includegraphics[scale = 0.8]{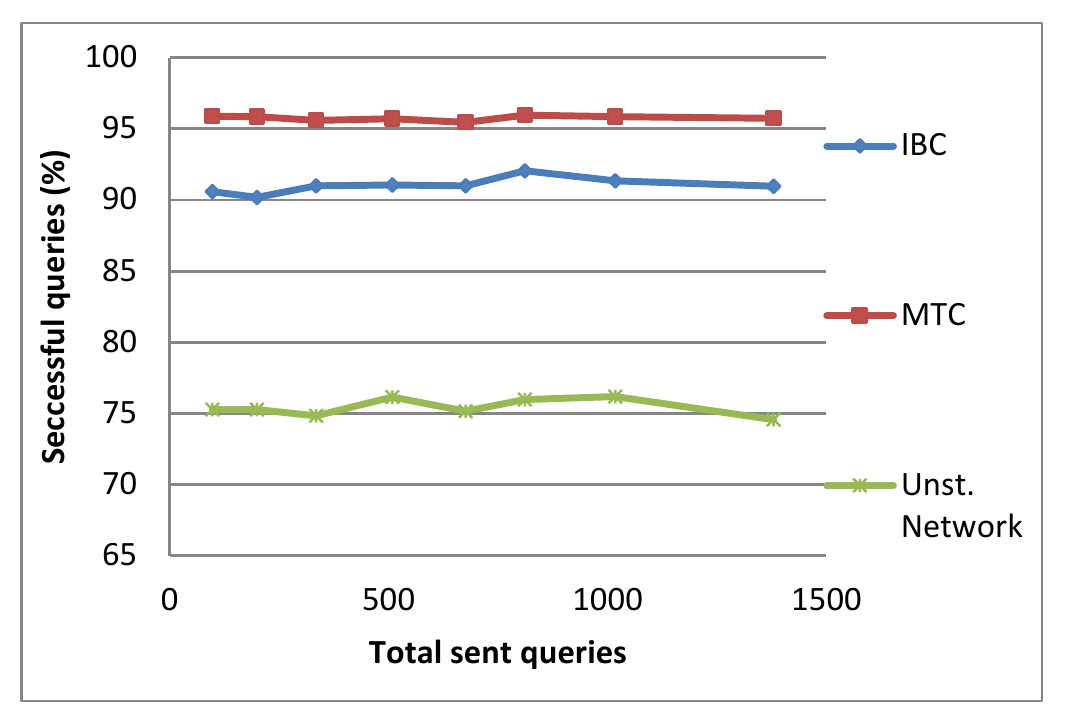}
	\caption{Successful queries as a result of increase in network traffic}
	\label{fig:fig12}
\end{figure}

The figures given below demonstrate the results obtained from our simulation. Fig.~\ref{fig:fig11} and Fig.~\ref{fig:fig12} show the rate of the successful queries sent in the network in the context of increasing network traffic and size. 

The unstructured network returns approximately 75\% successful queries out of the total queries sent in the network in both cases (i.e. for increasing network size and increasing network traffic). The success rate for IBCs is approximately 90\% of total queries sent in the network which is 20\% higher than for the unstructured network. The MTCs have approximately 95\% successful queries out of the total queries sent in the network, which represents a 26.67\% increase compared to the unstructured network and a 5.6\% increase compared to the IBCs. In order to better understand the effect on the network performance as a result of changes in size and network traffic, we calculated the Standard Deviation (SD) values for the rate of successful queries. Standard deviation indicates the variation of values from the mean value. The higher the value obtained from the SD, the higher the chances that a network is more affected by the size of and traffic in the network, which may make the network unstable during adverse conditions. The percentage values obtained were 1.086103055, 0.611178069 and 0.368911439 as a result of increasing the network size and 0.57384652, 0.50756941 and 0.1610037 as a result of increasing network traffic for the unstructured network, IBCs and MTCs respectively. The values for MTCs are lower than those for IBCs and unstructured networks in both cases which suggest it's likely to be more stable and can maintain the network performance longer than for the other two network structures. It's widely known that unstructured networks are not scalable and have poor performance as described above in Section~\ref{relwrk}, and the results obtained here corroborate this. Unstructured networks have difficulty handling increasing amounts of traffic, resulting in a higher SD value and noticeable deviation in the graphs in Fig.~\ref{fig:fig11} and Fig.~\ref{fig:fig12}. Performance starts to drop after a total of 1000 queries have been sent, as can clearly be seen in Fig.~\ref{fig:fig12}. The situation is similar for Fig.~\ref{fig:fig12}. The IBCs and MTCs successfully maintain their performance, but due to the instability of IBCs – as can be seen from the calculated SD values – we expect that performance will in general drop sooner than for MTCs. Also, based on the current results, we expect that the MTCs will maintain their performance for longer due to the higher query success rate and more stable underlying network structure.

\begin{figure}
\centering
	\includegraphics[scale = 0.8]{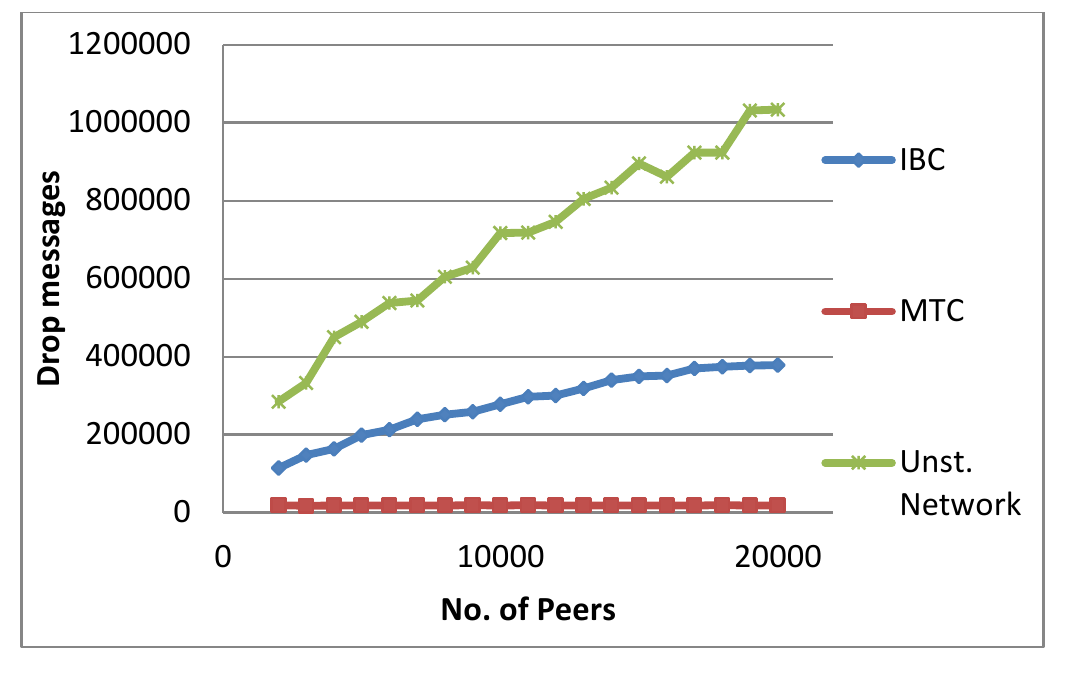}
	\caption{Network overhead produced by increase in network size}
	\label{fig:fig13}
\end{figure}

\begin{figure}
\centering
	\includegraphics[scale = 0.8]{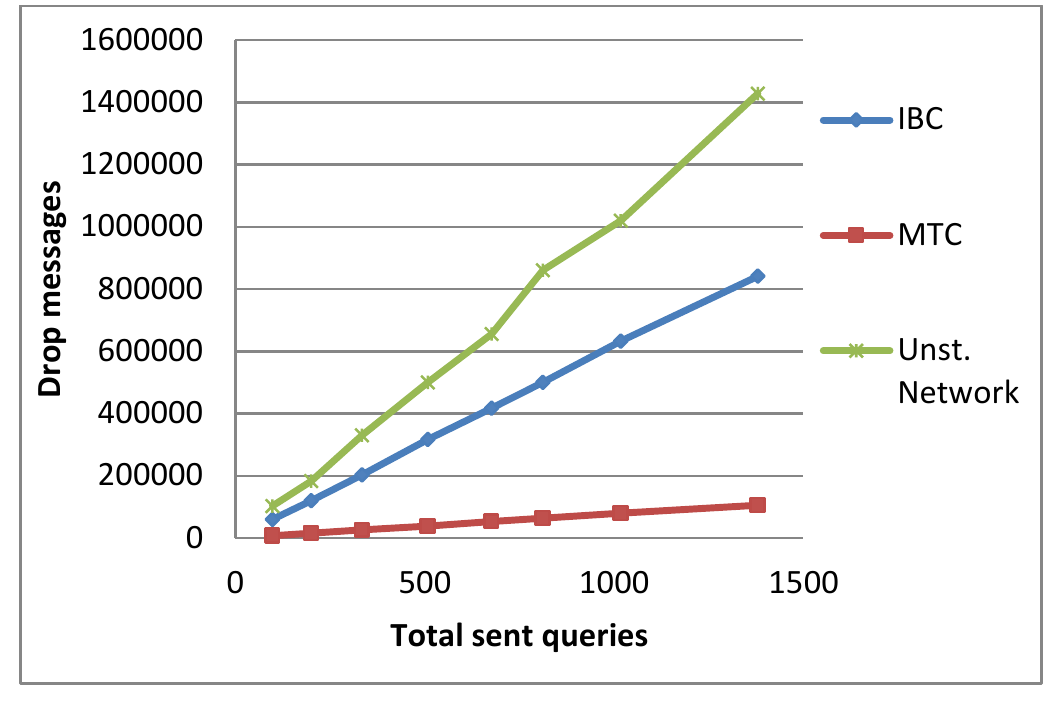}
	\caption{Network overhead produced by increase in network traffic}
	\label{fig:fig14}
\end{figure}

Interest-based communities reduce the network overhead as compared to unstructured networks, while the MTC structure reduces it further. The graphs in Fig.~\ref{fig:fig13} and Fig.~\ref{fig:fig14} show the overhead produced for each of these network structures. The network overhead is measured as the number of dropped messages during the experiment. A message is dropped when the Time-to-Live (TTL) of the message expires, a repetitive message is received or the peer that poses the query receives a drop message in response. The results show a considerable reduction in the number of dropped messages in the case of our MTCs as compared to IBCs and unstructured network. The MTC produces approximately seven and twelve times less network overhead than IBCs and 

\section{Conclusion and Future Work}
\label{conclu}

In this paper we presented the idea of memory threads to properly organize human digital memories and discussed how they can support social P2P networks. Memory threads are reflected in our Entity-based social P2P network as the underlying network structure in the form of Memory Thread-based Communities. The purpose of MTCs is to connect peers meaningfully. The data shared by a peer, which belongs to an entity, is used to connect it to other similar peers that are sharing data of the same entity. The connectivity inside an MTC is carried out according to an indexing criterion. This indexing criterion is a memory key, representing tagged information in the form of metadata produced as a result of data annotation. This allows the digital memories to be organised in a very effective way, as shown by our simulation results. The indexing criterion is reflected similarly in the MTC which organizes the peers according to the digital memories being shared. The resulting network improves performance, resulting in a higher rate of successful queries and lower network overhead as compared to existing network structures (unstructured networks and Interest-based communities).

The idea of Memory threads and MTCs is based on the idea that ``Let the data speaks for itself rather than represented by a peer or person'', which where lays the novelty of our work. Considering the importance of every digital memory and the details inside a digital memory of people, the idea of Entities in digital memories will consider every bit of a digital memory important and will be accessible in network through the MTCs. When further research is carried out and efficient ways are discovered to organize data, so our social P2P network will accommodate them and based on the results shown above improved performance of network is expected because network will be organized by the way people search for data.

We are currently building on this work by implementing sub-communities within the MTCs, which will be formed according to the indexing criteria of a memory thread to create a more robust network. As the above simulation is based on a static network, in the future we will also explore performance for MTCs implemented in dynamic scenarios, with peers frequently leaving and joining the network ('churn'). We are also implementing a search technique which operates on top of the MTCs, to determine the flexibility of the system with more complex search criteria. This may require the development of different structures for memory threads (not just linear) that can exist due to the different characteristics that appear in real world scenarios for various indexing values. The structure of a memory thread and the communities of peers which are formed according to the memory thread, depend upon the indexing criteria of the memory thread. The indexing criteria have a crucial effect on this structure. For example, while the history of an entity forms a linearly structured memory thread, the location at which data is captured might form a structure other than a linear structure.


%
%

\bibliographystyle{spmpsci}      
\bibliography{MyRef}


\end{document}